\documentclass[floatfix,pra,twocolumn,aps,superscriptaddress,showpacs]{revtex4-1}
\usepackage{hyperref}
\hypersetup{
	colorlinks=true,
	citecolor=blue,
	linkcolor=blue,
	urlcolor=blue}
\usepackage{graphicx}
\usepackage[normalem]{ulem}
\usepackage{amsmath}
\usepackage{amsfonts}
\usepackage{amssymb}
\usepackage{epsfig}
\usepackage{subfigure}
\usepackage{mathtools}
\usepackage{braket}
\usepackage[usenames,dvipsnames]{color}
\usepackage{setspace}
\usepackage{bm}
\usepackage{times}
\usepackage{ulem}
\usepackage{dsfont}
\usepackage{xcolor}

\begin{document}
\title{Temperature-induced supersolidity in spin-orbit-coupled Bose gases}
\author{Rajat}\email{rajat.19phz0009@iitrpr.ac.in}
\affiliation{Department of Physics, Indian Institute of Technology Ropar, Rupnagar-140001, Punjab, India}
\author{Ritu}\email{ritu.22phz0002@iitrpr.ac.in}
\affiliation{Department of Physics, Indian Institute of Technology Ropar, Rupnagar-140001, Punjab, India}
\author{Arko Roy}\email{arko@iitmandi.ac.in}
\affiliation{School of Physical Sciences, Indian Institute of Technology Mandi, Mandi-175075 (H.P.), India}
\author{Sandeep Gautam}\email{sandeep@iitrpr.ac.in}
\affiliation{Department of Physics, Indian Institute of Technology Ropar, Rupnagar-140001, Punjab, India}

\begin{abstract}
Close to the superfluid plane-wave (PW) - supersolid stripe (ST) phase transition point of a zero temperature quasi-one-dimensional 
spin-orbit-coupled Bose gas, we find that an increase in temperature induces a phase transition to the supersolid phase with a broken translational symmetry from the superfluid
plane-wave phase. We use the Hartree-Fock-Bogoliubov theory with the Popov approximation to investigate the effect of thermal fluctuations on the collective excitation spectrum 
and investigate the softening of the spin-dipole mode corresponding to the shift in the quantum critical point. This is in stark contrast to the PW-ST phase transition in a 
homogeneous system where non-zero temperatures facilitate the melting of the stripe phase.
\end{abstract}
\maketitle
\begin{figure}[!htbp]
\includegraphics[width=1.0\columnwidth]{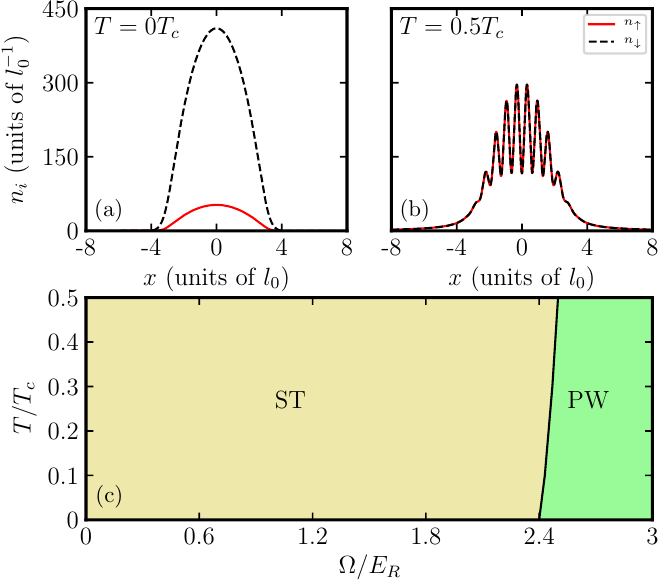}\\
\caption{Emergence of density modulation (supersolid phase) at non-zero temperature. 
The component total densities $n_{i}(x)$ for $\Omega = 2.45 E_R$ are shown in (a) at zero temperature $(T=0)$ 
illustrating the plane-wave (PW) phase, and in (b) at $T = 0.5 T_{c}$ illustrating the supersolid stripe (ST) phase. 
$\Omega_{\rm cr} = 2.4 E_R$ marks the critical transition point from PW to ST phase at $T=0$. 
(c) Finite-temperature phase diagram, in temperature $T$ versus Raman coupling strength $\Omega$ plane, 
illustrating the temperature-induced supersolidity in a trapped SO-coupled Bose gas. 
The indicator for such a transition through the softening of the spin-dipole mode, roton gap, and extraction of $T_c$ have been elaborated in the text.}
\label{PW-ST}
\end{figure}

Supersolid is a state of matter exhibiting simultaneously superfluid properties and 
 periodic spatial modulation in the particle density. 
With a quest to corroborate supersolidity in solid helium~\cite{PhysRevLett.109.155301} at ultracold temperatures, supersolid features have been demonstrated in the stripe phase 
of spin-orbit (SO) coupled spinor Bose-Einstein condensates (BECs) ~\cite{li2017stripe, PhysRevLett.124.053605}, dipolar Bose 
gases~\cite{PhysRevLett.122.130405, PhysRevX.9.011051, PhysRevX.9.021012, guo2019low, 
PhysRevLett.123.050402}, and BECs inside optical resonators~\cite{leonard2017supersolid}.
A Chester-type supersolid formed by the excitons in a semiconductor
double-layer heterostructure is also studied~\cite{PhysRevLett.130.057001}. The interplay of interaction between the atoms and the coupling strength induces a supersolid stripe (ST) phase
in SO-coupled pseudospin-1/2 BECs~\cite{PhysRevLett.105.160403, 
PhysRevLett.107.150403, PhysRevLett.108.225301, PhysRevLett.110.235302, PhysRevA.86.063621, 
Zheng_2013} at zero temperature, apart from the superfluid plane-wave (PW) and superfluid zero-momentum (ZM) phases. The ST phase ~\cite{PhysRevA.107.053302,  PhysRevLett.110.235302, 10.21468/SciPostPhys.11.5.092} is characterized by the two gapless Goldstone modes associated with the spontaneous breaking 
of $U(1)$ gauge and spatial translation symmetries~\cite{PhysRevLett.25.1543,li2017stripe, PhysRevLett.110.235302}. 
The transition from the ST to the PW phase is accompanied by the softening of the spin-dipole mode~\cite{PhysRevA.95.033616, PhysRevLett.127.115301,PhysRevLett.130.156001}. Non-zero temperatures are 
expected to shift the critical point, and previous works have confirmed that finite temperatures enhance the PW phase's domain in \emph{homogeneous} SO-coupled systems ~\cite{PhysRevA.106.023302, PhysRevA.96.013625, PhysRevA.90.053608}. That is, the supersolid melts and loses the ordered crystalline structure with the increase in temperature. 
This prediction, however, leaves out investigating the ST phase under external confinement, which we undertake in this paper.
For which, we calculate the excitation spectrum of a harmonically trapped quasi-one-dimensional (q1D) SO-coupled pseudo-spin-$1/2$ 
BEC at zero and finite temperatures using the Hartree-Fock-Bogoliubov (HFB) theory with the Popov approximation~\cite{PhysRevB.53.9341}. Our main results are shown in Fig.~\ref{PW-ST}, where
a non-zero temperature induces a phase transition from the superfluid PW to the supersolid ST phase, as demonstrated by the emergence of periodic density modulations. Additionally, by
detecting the minima of the spin-dipole mode at a shifted critical point, we confirm this temperature-induced supersolidity in the system. 
This is quite intriguing as, intuitively, one would expect the supersolid to melt with increased temperature. 

With an aim to study the effects of temperature on the quantum phases 
of SO-coupled BECs, we start with 
the dimensionless grand canonical Hamiltonian describing a q1D 
SO-coupled pseudo-spinor BEC extended along the $x$ direction in the second-quantized 
form~\cite{Lin2011}
\begin{align}\label{Ham}
\mathcal{H}=&\int dx \hat\psi_{i}^{\dagger}(x)\left[\left\{-\frac{1}{2}\frac{\partial^2}{\partial x^2}\ -\mu+V(x)\right\}\delta_{ij}\right.\nonumber\\
&-\iota k_{R} (\sigma_{z})_{ij}\frac{\partial}{\partial x}+\left. \frac{\Omega}{2} (\sigma_{x})_{ij}\right]\hat\psi_{j}(x) \nonumber\\
&+\frac{1}{2}\int  dx g_{ij} \hat\psi_{i}^{\dagger}(x) 
                      \hat\psi_{j}^{\dagger}(x) \hat\psi_{j}(x)    
                      \hat\psi_{i}(x),
\end{align}
where length, energy, and time are in units of $l_0 = \sqrt{\hbar/m\omega_x}$,
$\hbar\omega_x$, and $\omega_x^{-1}$, respectively, with $\omega_x$ as the harmonic
trapping angular frequency along the extended direction and $m$ as the atomic mass.
In Eq.~(\ref{Ham}), $\mu$ is the chemical potential, 
$V(x) = x^2/2$ is the trapping potential along the extended direction, 
recoil momentum $k_{R} = \sqrt{2 E_R}$ is a measure of the SO-coupling strength 
fixed by the recoil energy $E_R$ transferred to the atoms with spin indices $i, j \in (\uparrow,\downarrow)$ by the two counter-propagating laser beams, $\Omega$ is
the Raman coupling strength, $\sigma_{x,z}$ are the $2\times2$ Pauli spin$-1/2$ matrices,
and $g_{ij}$ denote the interaction strengths \cite{Lin2011}.
The renormalized interaction strengths in terms of the $s$-wave scattering lengths $a_{ij}$ are ${g}_{ij} = 2a_{ij}\omega_{\perp}/\omega_x$ with $\omega_{\perp}$ being the angular trapping frequency along the tightly confined $y$ and $z$ directions. 
To study the effects of quantum and thermal fluctuations on 
the supersolid phase, we extend the 
HFB theory using the Popov approximation~\cite{PhysRevB.53.9341, PhysRevA.89.013617, PhysRevA.106.013304} to the harmonically confined SO-coupled pseudo-spinor BEC.
Here we start with the Heisenberg equations of motion for the field operators, $\iota \partial \hat{\psi}_i/\partial t = [\hat{\psi}_i,\cal{H}]$, and split the Bose field operator into a mean-field
and a fluctuation operator, i.e. $\hat{\psi}_i =  \phi_i+\delta\hat{\psi}_i$. The ensemble average 
of the equations of motion (along with the mean-field approximation for the terms which are cubic
in fluctuation operators) yields the following time-independent coupled generalized Gross-Pitaevskii (GP) 
equations \cite{PhysRevA.106.013304}:
\begin{subequations}\label{gpe}
\begin{align}
\mu \phi_{\uparrow} =& \left[ -\frac{1}{2}\partial_x^2 + V + {g_{\uparrow\uparrow}( n_{\uparrow}+\tilde{n}_{\uparrow})+g_{\uparrow\downarrow}n_{\downarrow}}\right]\phi_{\uparrow}\nonumber\\&-\iota k_{R}\frac{\partial \phi_{\uparrow}}{\partial x}+\left(g_{\uparrow\downarrow} \tilde{n}_{\uparrow\downarrow}+\frac{\Omega}{2}\right) \phi_{\downarrow},\\
\mu \phi_{\downarrow} =& \left[ -\frac{1}{2}\partial_x^2 + V + {g_{\downarrow\downarrow}( n_{\downarrow}+\tilde{n}_{\downarrow})+g_{\uparrow\downarrow}n_{\uparrow}}\right]\phi_{\downarrow}\nonumber\\&+\iota k_{R}\frac{\partial \phi_{\downarrow}}{\partial x}+\left(g_{\uparrow\downarrow} \tilde{n}_{\downarrow\uparrow}+\frac{\Omega}{2}\right) \phi_{\uparrow},
\end{align}
\end{subequations}
where $ \tilde{n}_{ij} \equiv \langle\delta{\hat\psi}^{\dagger}_{i} \delta{\hat\psi}_{j}\rangle$, $n_i = |\phi_i|^2 + {\tilde n}_i$ is the total density of $i$th spin,
and the total number of atoms $N = \sum_{i} \int n_i dx$.
For simplicity of notations, we use ${\tilde n}_{i}$ instead of ${\tilde n}_{ii}$ 
to denote the thermal density of $i$th spin.

We subtract the generalized GP equations from the equations of motion for the field operators 
and then use the mean-field approximations $\delta{\hat\psi}_{i} \delta{\hat\psi}^{\dagger}_{j}\simeq\langle\delta{\hat\psi}_{i} \delta{\hat\psi}^{\dagger}_{j}\rangle$,
$\delta{\hat\psi}_{i} \delta{\hat\psi}_{j}\simeq\langle\delta{\hat\psi}_{i} \delta{\hat\psi}_{j}\rangle$ for the quadratic and  
$\delta{\hat\psi}^{\dagger}_{i} \delta{\hat\psi}_{j} \delta{\hat\psi}_{k} \simeq
\left\langle\delta{\hat\psi}^{\dagger}_{i} \delta{\hat\psi}_{j}\right\rangle 
\delta{\hat\psi}_{k}+\left\langle\delta{\hat\psi}^{\dagger}_{i} 
\delta{\hat\psi}_{k}\right\rangle \delta{\hat\psi}_{j}+\langle\delta{\hat\psi}_{j}
\delta{\hat\psi}_{k}\rangle \delta{\hat\psi}^{\dagger}_{i}$ for a typical cubic term
in the fluctuation operators \cite{PhysRevB.53.9341}. The anomalous average terms $\langle\delta{\hat\psi}_{i} \delta{\hat\psi}_{j}\rangle$ are negelected using Popov approximation to get the gapless spectrum~\cite{popov1987functional,PhysRevB.53.9341}. This leads to the equations of motion for the 
fluctuation operators, which, by the
Bogoliubov transformation, $\delta{\hat\psi}_{i}(x,t)=\sum_{\lambda}\left[u_{i}^{\lambda}(x) \hat{\alpha}_{\lambda}(x) 
e^{-i \omega_{\lambda} t}-v_{i}^{*\lambda}(x)\hat{\alpha}_{\lambda 
}^{\dagger}(x)e^{i\omega_{\lambda} t}\right]$, where ${\hat \alpha}^{\dagger}_\lambda ({\hat \alpha}_\lambda)$ is the quasi-particle creation (annihilation) operator,
lead to the Bogoliubov-de Gennes (BdG) equations for the quasi-particle amplitudes $u^{\lambda}$ and $v^{\lambda}$. These BdG equations are \cite{PhysRevB.53.9341, PhysRevA.106.013304}
\begin{equation}
\begin{pmatrix}
\mathcal{A} & -\mathcal{B}\\
\mathcal{B}^* & -\mathcal{A}^*
\end{pmatrix} 
\begin{pmatrix}
{\mathbf u}^{\lambda} \\ {\mathbf v}^{\lambda} 
\end{pmatrix} 
=\omega_{\lambda}
\begin{pmatrix}
{\mathbf u}^{\lambda} \\{\mathbf v}^{\lambda} 
\end{pmatrix},
\label{bdg}
\end{equation}
where
\begin{align*}
\mathcal{A}&=
\begin{pmatrix}
h_{0} + 2 g_{\uparrow\uparrow}n_{\uparrow}+g_{\uparrow\downarrow}{n}_{\downarrow} & {g_{\uparrow\downarrow}(\phi_{\downarrow}^* \phi_\uparrow+
\tilde{n}_{\uparrow\downarrow})+\frac{\Omega}{2}} \\
{g_{\uparrow\downarrow}(\phi_\uparrow^*\phi_{\downarrow}+\tilde{n}_{\downarrow\uparrow})+
\frac{\Omega}{2}} &  
h_{0} + 2 g_{\downarrow\downarrow}{n_\downarrow}+g_{\uparrow\downarrow}{n}_{\uparrow} \nonumber
\end{pmatrix},\\
\mathcal{B}&=
\begin{pmatrix}
{g_{\uparrow\uparrow}\phi_{\uparrow}^2} & 
{g_{\uparrow\downarrow}\phi_{\downarrow}\phi_{\uparrow}} \\
{g_{\uparrow\downarrow}\phi_{\uparrow}\phi_{\downarrow}}& 
g_{\downarrow\downarrow}\phi_{\downarrow}^{2}
\end{pmatrix},~{\mathbf u}^{\lambda} =\begin{pmatrix}
    u_{\uparrow}^{\lambda}\\ u_{\downarrow}^{\lambda}
\end{pmatrix},~ 
{\mathbf v}^{\lambda}= \begin{pmatrix}
    v_{\uparrow}^{\lambda}\\ v_{\downarrow}^{\lambda}
\end{pmatrix}.
\end{align*}
Here $h_0 = - {\frac {1}{2}}\partial_x^2 \mp\iota k_{R} \partial_x + V - \mu$, $  \tilde{n}_{ij} \equiv \langle\delta{\hat\psi}^{\dagger}_{i} \delta{\hat\psi}_{j}\rangle= \sum_{\lambda}\left\{\left(u_{i}^{\lambda *} u_{j}^{\lambda} +v_{i}^{\lambda} v_{j}^{\lambda *}\right)
  f_{\omega_{\lambda}} +v_{i}^{\lambda} v_{j}^{\lambda *}\right\}$ with $f_{\omega_\lambda}$ as the Bose-Einstein distribution function.
 The $\lambda$th quasiparticle amplitudes are 
 normalized as $\int{\sum_{i}(|u_{i}^{\lambda}|^{2}-|v_{i}^{\lambda}|^{2})}dx = 1$. 
 Eqs.~(\ref{gpe}) and (\ref{bdg}) are to be solved self-consistently to
 calculate finite-temperature densities and excitation spectrum. 
In the first step of the self-consistent iterative calculation, 
we numerically solve Eqs.~(\ref{gpe}a) and (\ref{gpe}b) to obtain $\phi_{\uparrow} (\phi_{\downarrow})$ using split time-step imaginary-time method~\cite{kaur2021fortress,*banger2021semi,*banger2022fortress}.
These condensate
wavefunctions are input to Eq.~(\ref{bdg}) in the second step. With the quasiparticle, amplitudes expanded in terms of the harmonic oscillator eigenfunctions, Eq.~(\ref{bdg}) yields a generalized matrix eigenvalue problem for the expansion coefficients~\cite{PhysRevA.57.R32,roy-2020}. We solve the eigenvalue problem using the standard matrix diagonalization algorithms to obtain the 
eigenenergies ($\omega_{\lambda}$) and quasiparticle amplitudes ($u^{\lambda}$ and $v^{\lambda}$ ),
which are used to evaluate (updated) $\tilde {n}_{ij}$ and
renormalize the condensate wavefunctions to ensure $ \int (|\phi_{\uparrow}|^2+|\phi_{\downarrow}|^2)dx= N -  \sum_i \int \tilde{n}_{i} dx$. 
The $\tilde {n}_{ij}$ and the updated $\phi_{\uparrow} (\phi_{\downarrow})$ are fed back to Eqs.~(\ref{gpe}a) and (\ref{gpe}b) to repeat the previous two steps.
This procedure is executed iteratively until thermal densities converge within a chosen tolerance limit~\cite{PhysRevA.106.013304, PhysRevA.57.R32,roy-2020}. We consider a spatial grid consisting of $4096$ points with a spatial step size of $\Delta x = 0.01$ and imaginary-time step of $10^{-5}$ to solve Eqs. (\ref{gpe}a)-(\ref{gpe}b), and $200$ harmonic oscillator basis states to solve
BdG equation (\ref{bdg}). We have checked that increasing the size of the basis does
not affect the results.


With these definitions and considerations, we consider a pseudo-spinor system consisting of two hyperfine spin states
of $^{23}$Na in a highly anisotropic harmonic trap with $\omega_x = 2\pi\times5$ Hz, 
$\omega_{\perp}=\omega_y = \omega_z = 20 \omega_x$ such that  $\omega_{\perp} \gg 
\omega_{x}$, and $N = 2000$ at $T=0$.
In this case, we can integrate out the $y$ and $z$ coordinates from 
the condensate wave function and describe the system as a q1D system along the $x$ axis.
We consider $a_{\uparrow\uparrow} = a_{\downarrow\downarrow} = 54.54 a_0/l_0$ as
the (dimensionless) intra-spin $s$-wave scattering length, where $a_0$ is Bohr radius \cite{PhysRevA.83.042704}.
To enhance the visibility of the stripe phase, we consider reduced interspecies interactions. This may be achieved by reducing the spatial overlap
between the wave functions of the two spin components. One way to accomplish this is by employing a spin-dependent trapping potential separating the two components~\cite{PhysRevA.90.041604}. Another method is to utilize pseudo-spin orbital states within a superlattice potential, which has already been implemented in an experiment with spinor BEC of $^{23}$Na.~\cite{li2017stripe}.
Given these we consider $g_{\uparrow\downarrow} = g_{\downarrow\uparrow} = 0.6 g_{\uparrow\uparrow}$, where  $g_{\uparrow\uparrow} = 2a_{\uparrow\uparrow}\omega_{\perp}/\omega_x$.
We will use the pseudo-spinor BEC with these parameters ($N$, $g_{\uparrow \uparrow}$, $g_{\uparrow\downarrow}$, $\omega_x$, $\omega_{\perp}$, and $k_R$) throughout this paper, except if specifically stated otherwise.
To investigate the effects of temperature on phase transition, at the outset, 
we vary the Raman coupling strength $\Omega$ while keeping the SO-coupling strength fixed at $k_R=6$ 
in this work. At $T=0$, with an increase in $\Omega$, the system first undergoes a phase transition from 
the ST to the PW phase at a critical Raman coupling strength $\Omega_{\rm cr} \approx 2.4 E_R$, 
which is followed by a second phase transition to the ZM phase~\cite{PhysRevLett.107.150403, PhysRevLett.108.225301}.
As ST and PW phases have qualitatively distinct density profiles, 
we first examine the total densities (condensate plus thermal density) of the two spin states obtained by
solving Eqs.~(\ref{gpe}a)-(\ref{gpe}b) and  (\ref{bdg}).
In Figs.~\ref{PW-ST}(a) and \ref{PW-ST}(b), we present the total density 
$n_{i}$ for $\Omega = 2.45 E_R$ at two different temperatures: $T = 0$ and $T = 0.5 T_{c}$, where
$T_c$ is the critical temperature for the SO-coupled BEC calculated using the HFB-Popov method and discussed in the following paragraph. 
At zero temperature, the densities in Fig.~\ref{PW-ST}(a) correspond to the PW phase (as $\Omega > 2.4E_R$) with a non-zero magnetization [$\int\{ 
n_{\uparrow}(x) - n_{\downarrow}(x)\}dx$ ]. Surprisingly, at $T = 0.5 T_{c}$
in Fig.~\ref{PW-ST}(b),
the densities $n_i$ exhibit distinctive modulations of the ST phase
with zero magnetization, indicating that the system is in the stripe phase. 
This suggests a shift in the ST-PW phase boundary with temperature [cf. Fig.~\ref{PW-ST}(c)].
Furthermore, we calculate the excitation spectrum of the system to determine the phase boundary between the two phases at different temperatures. In the ST phase of a harmonically trapped BEC, 
the spin-dipole mode softens with an increase in $\Omega$ and acquires a minimum at the 
ST-PW phase boundary. Across the same phase boundary, there is a slight jump in the dipole and breathing modes, indicating a first-order transition between supersolid and superfluid states~\cite{PhysRevA.95.033616, PhysRevLett.127.115301}. 

\begin{figure}[!htbp]
\includegraphics[width=\columnwidth]{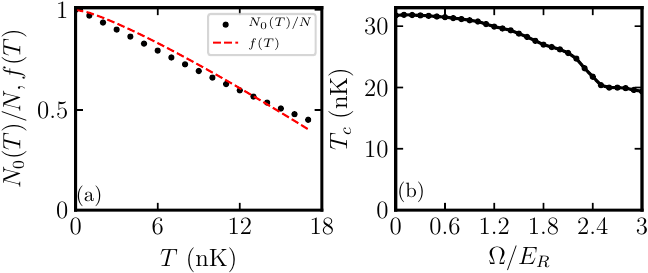}
\caption{(a) shows the plot of the condensate fraction $N_0(T)/N$
obtained from the HFB-Popov theory for the pseudo-spinor BEC with $\Omega = 2$ 
and the fitting function $f(T)$, where 
$N_0(T)/N \geqslant 0.45$ (criterion for the applicability of the HFB-Popov theory as discussed in the main text), 
$f(T) = 1-T \ln (2 k_B T /\hbar \omega_x) / T_c \ln \left(2 k_B T_c / \hbar \omega_x\right)$ with $T_c (= 26.21~{\rm nK})$ as the fitting parameter. (b) Variation of $T_c$ with $\Omega/E_R$.}
\label{con_frac}
\end{figure}

Before elaborating on the role of the excitation spectrum at non-zero temperatures indicating the modified ST-PW phase boundary, it is worthwhile to outline the procedure involved in the computation of $T_c$  and illustrate the reliability of the present theoretical framework by ascertaining the domain of applicability of the HFB Popov theory.
For which, first, using the HFB-Popov theory, we calculate the condensate fraction $N_0(T)/N$ of the
pseudo-spinor BEC as a function of temperature $T$, but without SO and Raman couplings, i.e.
$k_R = \Omega = 0$, by solving Eqs.~(\ref{gpe}a), (\ref{gpe}b), and (\ref{bdg}) self-consistently 
with $N_0 = \sum_i \int |\phi_i|^2 dx$ being the total number of atoms in the condensate. 
We then extract the critical temperature $T_c$ by 
fitting the condensate fraction data points using the function $f(T) = 1-T \ln (2 k_B T /\hbar \omega_x) / T_c \ln \left(2 k_B T_c / \hbar \omega_x\right)$ (analytic estimate for the condensate fraction)~\cite{PhysRevA.54.656} with $T_c$ as the fitting parameter. We observe that for 
$N_0(T)/N \approx 0.45$ as obtained from the HFB-Popov theory, we find an excellent agreement of the extracted $T_c$ with the analytical prediction $T_c^{\rm ana.} = \hbar \omega_{x} N/ (2 k_B\ln N)$~\cite{PhysRevA.54.656}. We set this value of $N_0(T)/N \geqslant 0.45$ as the benchmark for the validity of the HFB-Popov theory and use it to compute $T_c$ in the presence of SO and Raman couplings. In Fig.~\ref{con_frac}(a), we plot the fitting function $f(T)$ and $N_0(T)/N$ for the pseudo-spinor BEC with $\Omega = 2$ as a demonstrative example. As expected, the critical temperature varies with $\Omega$ as shown in Fig.~\ref{con_frac}(b)~\cite{ji2014experimental,PhysRevA.89.063614}.
\begin{figure}[!htbp]
\includegraphics[width=\columnwidth]{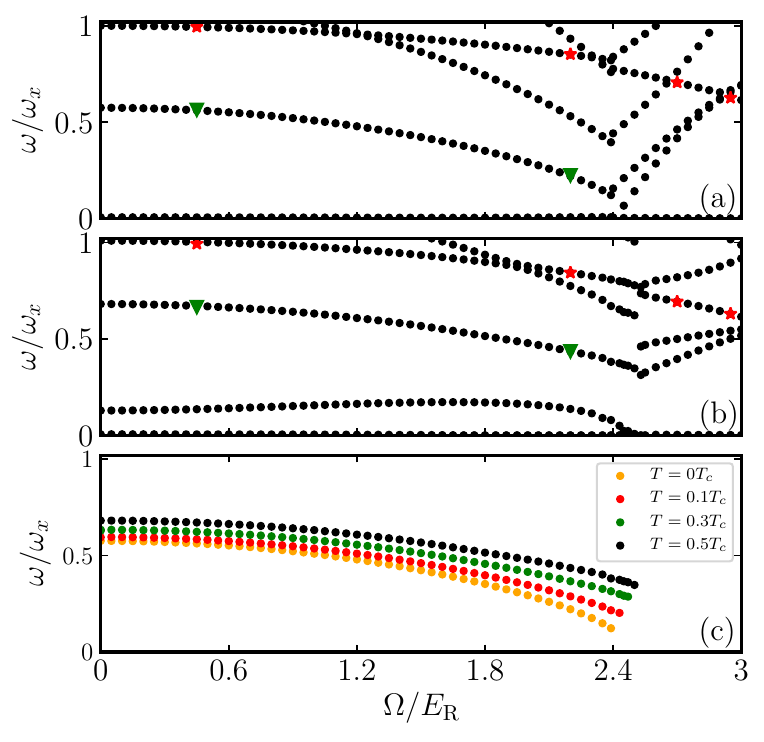}
\caption{The excitation spectrum of the harmonically trapped SO-coupled q1D BEC at (a) $T=0$ and (b)  $T=0.5 T_c(\Omega)$. The dark green triangles are the markers for the spin-dipole
mode, which gets softened as one approaches the ST-PW phase boundary at a critical coupling strength of $\Omega_{\rm cr} \approx 2.4 E_R$ in (a) and $\Omega_{\rm cr} \approx 2.5 E_R$ in (b). The light red stars mark the dipole mode, which shows a discontinuity at the ST-PW phase boundary, reflecting the first-order nature of the transition. 
(c) The softening of spin dipole modes with the $\Omega$ for different temperatures.
For a fixed $\Omega/E_R$, the energy of the spin-dipole mode increases with temperature in the ST phase.}
\label{ft}
\end{figure}

In Fig.~\ref{ft}(a), we present the variation in the excitation frequencies at zero temperature with the increase in the Raman coupling strength 
$\Omega$. We focus on the spin-dipole and density-dipole modes to map out the phase boundary~\cite{PhysRevA.95.033616, PhysRevLett.127.115301}. 
The dipole mode (marked by the light red star) in the limit $\Omega \rightarrow 0$ is $\omega_D = \omega_x$, satisfying the Kohn theorem. The spin-dipole mode (marked by the dark green triangle) in Fig.~\ref{ft}(a) decreases as $\Omega$ increases and acquires a minimum value at the ST-PW phase boundary ($\Omega_{\rm cr} \approx 2.4 E_R$). 
Furthermore, the spin-dipole mode is identified by extracting the oscillation frequency of $\int x [n_{\uparrow}(x,t) - n_{\downarrow}(x,t)]dx$ during the time evolution of the ST 
ground-state phase governed by $T = 0$ GP equations corresponding to the perturbed Hamiltonian \cite{PhysRevA.108.043310}, where perturbation $\propto x \sigma_z$.
The density-dipole mode in Fig.~\ref{ft}(a) shows a discontinuous change at the ST-PW phase boundary, reflecting the first-order nature of the transition~\cite{PhysRevLett.127.115301, PhysRevLett.130.156001}. The dipole mode is similarly identified by extracting the oscillation frequency of $ \int x [n_{\uparrow}(x,t) + n_{\downarrow}(x,t)] dx$ after a perturbation  $\propto x$ is added to the Hamiltonian. It is to be noted that the ST phase has two Goldstone modes: the density-Goldstone mode due to the $U(1)$ gauge-symmetry breaking and the 
spin-Goldstone mode due to the breaking of translational invariance symmetry. We obtained two Goldstone modes from our calculations, which serve as a consistency check of our numerical calculations. 
In the PW phase ($\Omega>\Omega_{\rm cr}$), a single density-Goldstone mode is identified corresponding to the $U(1)$ 
gauge-symmetry breaking. In this phase, as bosons condense into one of the minima of the single-particle dispersion, it breaks the discrete $\mathbb{Z}_2$ symmetry, acquiring non-zero magnetization ~\cite{Zheng_2013}. The spin-dipole mode is absent in this phase, whereas the non-zero density-dipole
mode decreases with increasing $\Omega$.

Fig.~\ref{ft}(b) shows the excitation spectrum at $T = 0.5T_c (\Omega)$, where, like $T = 0$, the ST-PW phase boundary is characterized by the minima 
of the spin-dipole mode in the ST phase and a discontinuous change 
in the density-dipole mode across the phase boundary. The phase boundary 
has shifted to the right with $\Omega_{\rm cr} \approx 2.5E_R$, which
is consistent with the ST phase in Fig.~\ref{PW-ST}(b). It is to be noted that the condensate fraction $N_0(T)/N \approx 0.55$ for the range of $\Omega$ considered. Furthermore, the spin-Goldstone mode, nearly zero at $T=0$ in Fig.~\ref{ft}(a), is now non-zero, possibly due to 
the thermal effects.
To see the shifting of phase boundary with temperature distinctly, we calculate spin-dipole
modes at $T=0$, $0.1 T_{c}$, $0.3 T_{c}$, and $0.5 T_{c}$, shown in
Fig.~\ref{ft}(c). The $\Omega_{\rm cr}$ increases with an increase
in temperature, resulting in the ST phase expanding into the domain of the
PW phase. This temperature-induced enhancement of the ST phase in q1D condensates, and the emergence of supersolidity is the key result of this work.
On the contrary, the ST phase shrinks with temperature for a three-dimensional SO-coupled homogeneous pseudo-spinor BEC~\cite{PhysRevA.96.013625, PhysRevA.90.053608}. To substantiate this, we compute 
the dispersion to monitor the variation of the roton gap in the homogeneous and the q1D trapped spinor system with similar parameters. It is to be noted that for the trapped BEC, we arrive at the dispersion relation using the Fourier transforms of the Bogoliubov quasiparticle amplitudes $u_{i}^{\lambda}(x)$ and 
$v_{i}^{\lambda}(x)$, and then calculate the root-mean-square wave number $k_{\rm rms}$ of the $\lambda$th quasiparticle mode 
as given in Ref.~\cite{PhysRevLett.104.094501}. At $T=0$, the phase transition from the ST to the PW phase is marked by the opening of the roton gap ($\Delta$) at a finite wavenumber $k$~\cite{PhysRevA.90.063624, PhysRevLett.114.105301}.
\begin{figure}[!htbp]
\includegraphics[width=\columnwidth]{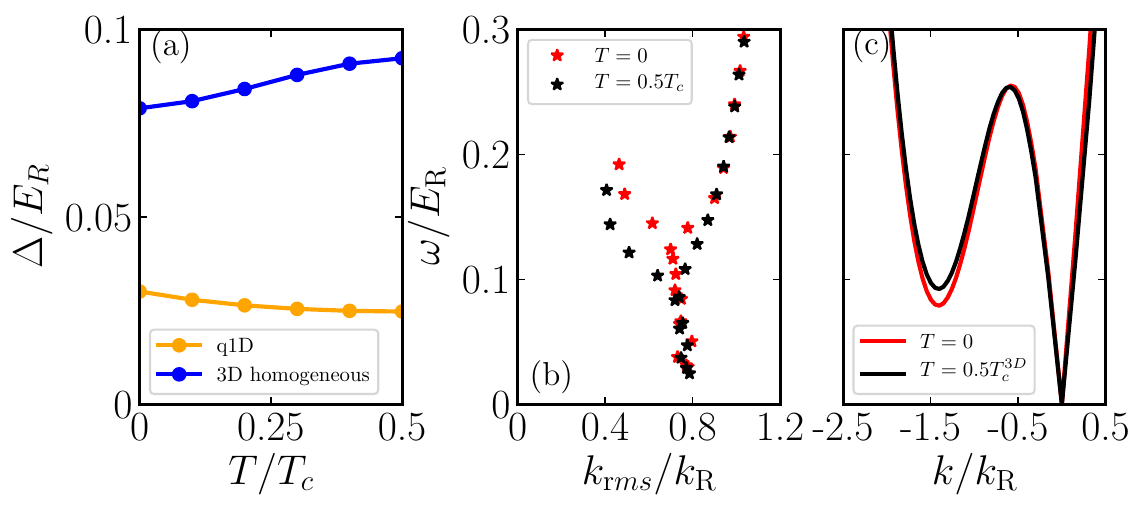}
\caption{(a) shows the increase (decrease) of the roton gap $\Delta$ in the SO-coupled homogeneous (q1D trapped pseudo-spinor) BEC with temperature. (b) Dispersion of the trapped pseudo-spinor BEC's at temperature $T = 0.5 T_{c}$ revealing roton fingers~\cite{PhysRevA.88.043606}. (c) shows the dispersion for the homogeneous SO-coupled pseudo-spinor BEC. For the homogeneous system, the uniform density considered in (a) and (c) equals the peak density [$n(x=0) = 467.46$] of the trapped SO-coupled BEC. This gives $g_{\uparrow\uparrow} n = 0.48034 E_R$ and $g_{\uparrow\downarrow} n = 0.6 g_{\uparrow\uparrow} n $ for the homogeneous SO-coupled BEC. $T_c^{3D}= 2 \pi \hbar^2 [(n/2)/\zeta(3/2)]^{2/3}/m k_{B}$ is the critical temperature of ideal three-dimensional two-component Bose gas. In (a)-(c), Raman coupling strength $\Omega = 2.8 E_R$. Contrary to the trapped SO-coupled BEC, the roton mode increases with temperature for the homogeneous SO-coupled Bose gas.}
\label{disp}
\end{figure}
We use this measure as an additional indicator to emphasize that the domain of the ST phase indeed increases at finite temperatures for the system considered in this work. As illustrated in Fig.~\ref{disp}(a), we find that the roton gap increases (decreases) in the homogeneous (harmonically-trapped q1D pseudo-spinor BEC) with temperature. The increase in the gap confirms the melting of the ST phase into the PW phase with temperature ~\cite{PhysRevA.96.013625, PhysRevA.90.053608},  whereas the decrease affirms the emergence of the ST phase with temperature in the q1D system.
In Figs.~\ref{disp}(b) and~\ref{disp}(c), we have shown the dispersion curve for the q1D trapped and homogeneous SO-coupled BEC at zero and a finite temperature of $T = 0.5 T_c$.
For the trapped SO-coupled system, Fig.~\ref{disp}(b) reveals the so-called roton finger~\cite{PhysRevA.88.043606}.

These observations strengthen the fact that at non-zero temperatures, the PW phase undergoes a transition to the supersolid phase identified by density stripes. By using the Hartree-Fock-Bogoliubov 
theory with the Popov approximation and including the interactions between the thermal atoms of the different components, we have demonstrated the softening of the spin-dipole mode accompanied by a 
shift of the quantum critical point. The domain of the PW phase, characterized by a finite roton gap ($\Delta$), decreases with increasing temperature and is consistent with the emergence of 
supersolidity at finite temperatures. Possible areas for future research include the effects of dimensionality and finite-size scaling of the quantum-critical point. Recently, we became aware of such 
temperature-induced transition in a dipolar BEC, where non-zero temperatures induce a transition from a dipolar superfluid to a supersolid state~\cite{PhysRevLett.126.233401, sanchez2023heating}. This 
suggests that temperature-induced supersolidity manifests the semblance between the dipolar and SO-coupled supersolids. Recently, a generic mechanism for a thermally-driven reentrant supersolidity has also been proposed~\cite{radzihovsky2023reentrant}.

A.R. acknowledges the support of the Science and Engineering Research Board
(SERB), Department of Science and Technology, Government of India, under the project SRG/2022/000057
and IIT Mandi seed-grant funds under the project
IITM/SG/AR/87. A.R. acknowledges the National Supercomputing Mission (NSM) for providing computing resources of PARAM Himalaya at IIT Mandi,
which is implemented by C-DAC and supported by the Ministry of Electronics and Information Technology (MeitY) and Department of Science
and Technology (DST), Government of India. S.G. acknowledges support from the Science and Engineering Research Board, Department
of Science and Technology, Government of India through Project No. CRG/2021/002597.
\bibliography{bib_file}{}

\begin{thebibliography}{48}%
\makeatletter
\providecommand \@ifxundefined [1]{%
 \@ifx{#1\undefined}
}%
\providecommand \@ifnum [1]{%
 \ifnum #1\expandafter \@firstoftwo
 \else \expandafter \@secondoftwo
 \fi
}%
\providecommand \@ifx [1]{%
 \ifx #1\expandafter \@firstoftwo
 \else \expandafter \@secondoftwo
 \fi
}%
\providecommand \natexlab [1]{#1}%
\providecommand \enquote  [1]{``#1''}%
\providecommand \bibnamefont  [1]{#1}%
\providecommand \bibfnamefont [1]{#1}%
\providecommand \citenamefont [1]{#1}%
\providecommand \href@noop [0]{\@secondoftwo}%
\providecommand \href [0]{\begingroup \@sanitize@url \@href}%
\providecommand \@href[1]{\@@startlink{#1}\@@href}%
\providecommand \@@href[1]{\endgroup#1\@@endlink}%
\providecommand \@sanitize@url [0]{\catcode `\\12\catcode `\$12\catcode
  `\&12\catcode `\#12\catcode `\^12\catcode `\_12\catcode `\%12\relax}%
\providecommand \@@startlink[1]{}%
\providecommand \@@endlink[0]{}%
\providecommand \url  [0]{\begingroup\@sanitize@url \@url }%
\providecommand \@url [1]{\endgroup\@href {#1}{\urlprefix }}%
\providecommand \urlprefix  [0]{URL }%
\providecommand \Eprint [0]{\href }%
\providecommand \doibase [0]{http://dx.doi.org/}%
\providecommand \selectlanguage [0]{\@gobble}%
\providecommand \bibinfo  [0]{\@secondoftwo}%
\providecommand \bibfield  [0]{\@secondoftwo}%
\providecommand \translation [1]{[#1]}%
\providecommand \BibitemOpen [0]{}%
\providecommand \bibitemStop [0]{}%
\providecommand \bibitemNoStop [0]{.\EOS\space}%
\providecommand \EOS [0]{\spacefactor3000\relax}%
\providecommand \BibitemShut  [1]{\csname bibitem#1\endcsname}%
\let\auto@bib@innerbib\@empty
\bibitem [{\citenamefont {Kim}\ and\ \citenamefont
  {Chan}(2012)}]{PhysRevLett.109.155301}%
  \BibitemOpen
  \bibfield  {author} {\bibinfo {author} {\bibfnamefont {D.~Y.}\ \bibnamefont
  {Kim}}\ and\ \bibinfo {author} {\bibfnamefont {M.~H.~W.}\ \bibnamefont
  {Chan}},\ }\href {\doibase 10.1103/PhysRevLett.109.155301} {\bibfield
  {journal} {\bibinfo  {journal} {Phys. Rev. Lett.}\ }\textbf {\bibinfo
  {volume} {109}},\ \bibinfo {pages} {155301} (\bibinfo {year}
  {2012})}\BibitemShut {NoStop}%
\bibitem [{\citenamefont {Li}\ \emph {et~al.}(2017)\citenamefont {Li},
  \citenamefont {Lee}, \citenamefont {Huang}, \citenamefont {Burchesky},
  \citenamefont {Shteynas}, \citenamefont {Top}, \citenamefont {Jamison},\ and\
  \citenamefont {Ketterle}}]{li2017stripe}%
  \BibitemOpen
  \bibfield  {author} {\bibinfo {author} {\bibfnamefont {J.-R.}\ \bibnamefont
  {Li}}, \bibinfo {author} {\bibfnamefont {J.}~\bibnamefont {Lee}}, \bibinfo
  {author} {\bibfnamefont {W.}~\bibnamefont {Huang}}, \bibinfo {author}
  {\bibfnamefont {S.}~\bibnamefont {Burchesky}}, \bibinfo {author}
  {\bibfnamefont {B.}~\bibnamefont {Shteynas}}, \bibinfo {author}
  {\bibfnamefont {F.~{\c{C}}.}\ \bibnamefont {Top}}, \bibinfo {author}
  {\bibfnamefont {A.~O.}\ \bibnamefont {Jamison}}, \ and\ \bibinfo {author}
  {\bibfnamefont {W.}~\bibnamefont {Ketterle}},\ }\href {\doibase
  10.1038/nature21431} {\bibfield  {journal} {\bibinfo  {journal} {Nature}\
  }\textbf {\bibinfo {volume} {543}},\ \bibinfo {pages} {91} (\bibinfo {year}
  {2017})}\BibitemShut {NoStop}%
\bibitem [{\citenamefont {Putra}\ \emph {et~al.}(2020)\citenamefont {Putra},
  \citenamefont {Salces-C\'arcoba}, \citenamefont {Yue}, \citenamefont
  {Sugawa},\ and\ \citenamefont {Spielman}}]{PhysRevLett.124.053605}%
  \BibitemOpen
  \bibfield  {author} {\bibinfo {author} {\bibfnamefont {A.}~\bibnamefont
  {Putra}}, \bibinfo {author} {\bibfnamefont {F.}~\bibnamefont
  {Salces-C\'arcoba}}, \bibinfo {author} {\bibfnamefont {Y.}~\bibnamefont
  {Yue}}, \bibinfo {author} {\bibfnamefont {S.}~\bibnamefont {Sugawa}}, \ and\
  \bibinfo {author} {\bibfnamefont {I.~B.}\ \bibnamefont {Spielman}},\ }\href
  {\doibase 10.1103/PhysRevLett.124.053605} {\bibfield  {journal} {\bibinfo
  {journal} {Phys. Rev. Lett.}\ }\textbf {\bibinfo {volume} {124}},\ \bibinfo
  {pages} {053605} (\bibinfo {year} {2020})}\BibitemShut {NoStop}%
\bibitem [{\citenamefont {Tanzi}\ \emph {et~al.}(2019)\citenamefont {Tanzi},
  \citenamefont {Lucioni}, \citenamefont {Fam\`a}, \citenamefont {Catani},
  \citenamefont {Fioretti}, \citenamefont {Gabbanini}, \citenamefont {Bisset},
  \citenamefont {Santos},\ and\ \citenamefont
  {Modugno}}]{PhysRevLett.122.130405}%
  \BibitemOpen
  \bibfield  {author} {\bibinfo {author} {\bibfnamefont {L.}~\bibnamefont
  {Tanzi}}, \bibinfo {author} {\bibfnamefont {E.}~\bibnamefont {Lucioni}},
  \bibinfo {author} {\bibfnamefont {F.}~\bibnamefont {Fam\`a}}, \bibinfo
  {author} {\bibfnamefont {J.}~\bibnamefont {Catani}}, \bibinfo {author}
  {\bibfnamefont {A.}~\bibnamefont {Fioretti}}, \bibinfo {author}
  {\bibfnamefont {C.}~\bibnamefont {Gabbanini}}, \bibinfo {author}
  {\bibfnamefont {R.~N.}\ \bibnamefont {Bisset}}, \bibinfo {author}
  {\bibfnamefont {L.}~\bibnamefont {Santos}}, \ and\ \bibinfo {author}
  {\bibfnamefont {G.}~\bibnamefont {Modugno}},\ }\href {\doibase
  10.1103/PhysRevLett.122.130405} {\bibfield  {journal} {\bibinfo  {journal}
  {Phys. Rev. Lett.}\ }\textbf {\bibinfo {volume} {122}},\ \bibinfo {pages}
  {130405} (\bibinfo {year} {2019})}\BibitemShut {NoStop}%
\bibitem [{\citenamefont {B\"ottcher}\ \emph {et~al.}(2019)\citenamefont
  {B\"ottcher}, \citenamefont {Schmidt}, \citenamefont {Wenzel}, \citenamefont
  {Hertkorn}, \citenamefont {Guo}, \citenamefont {Langen},\ and\ \citenamefont
  {Pfau}}]{PhysRevX.9.011051}%
  \BibitemOpen
  \bibfield  {author} {\bibinfo {author} {\bibfnamefont {F.}~\bibnamefont
  {B\"ottcher}}, \bibinfo {author} {\bibfnamefont {J.-N.}\ \bibnamefont
  {Schmidt}}, \bibinfo {author} {\bibfnamefont {M.}~\bibnamefont {Wenzel}},
  \bibinfo {author} {\bibfnamefont {J.}~\bibnamefont {Hertkorn}}, \bibinfo
  {author} {\bibfnamefont {M.}~\bibnamefont {Guo}}, \bibinfo {author}
  {\bibfnamefont {T.}~\bibnamefont {Langen}}, \ and\ \bibinfo {author}
  {\bibfnamefont {T.}~\bibnamefont {Pfau}},\ }\href {\doibase
  10.1103/PhysRevX.9.011051} {\bibfield  {journal} {\bibinfo  {journal} {Phys.
  Rev. X}\ }\textbf {\bibinfo {volume} {9}},\ \bibinfo {pages} {011051}
  (\bibinfo {year} {2019})}\BibitemShut {NoStop}%
\bibitem [{\citenamefont {Chomaz}\ \emph {et~al.}(2019)\citenamefont {Chomaz},
  \citenamefont {Petter}, \citenamefont {Ilzh\"ofer}, \citenamefont {Natale},
  \citenamefont {Trautmann}, \citenamefont {Politi}, \citenamefont
  {Durastante}, \citenamefont {van Bijnen}, \citenamefont {Patscheider},
  \citenamefont {Sohmen}, \citenamefont {Mark},\ and\ \citenamefont
  {Ferlaino}}]{PhysRevX.9.021012}%
  \BibitemOpen
  \bibfield  {author} {\bibinfo {author} {\bibfnamefont {L.}~\bibnamefont
  {Chomaz}}, \bibinfo {author} {\bibfnamefont {D.}~\bibnamefont {Petter}},
  \bibinfo {author} {\bibfnamefont {P.}~\bibnamefont {Ilzh\"ofer}}, \bibinfo
  {author} {\bibfnamefont {G.}~\bibnamefont {Natale}}, \bibinfo {author}
  {\bibfnamefont {A.}~\bibnamefont {Trautmann}}, \bibinfo {author}
  {\bibfnamefont {C.}~\bibnamefont {Politi}}, \bibinfo {author} {\bibfnamefont
  {G.}~\bibnamefont {Durastante}}, \bibinfo {author} {\bibfnamefont {R.~M.~W.}\
  \bibnamefont {van Bijnen}}, \bibinfo {author} {\bibfnamefont
  {A.}~\bibnamefont {Patscheider}}, \bibinfo {author} {\bibfnamefont
  {M.}~\bibnamefont {Sohmen}}, \bibinfo {author} {\bibfnamefont {M.~J.}\
  \bibnamefont {Mark}}, \ and\ \bibinfo {author} {\bibfnamefont
  {F.}~\bibnamefont {Ferlaino}},\ }\href {\doibase 10.1103/PhysRevX.9.021012}
  {\bibfield  {journal} {\bibinfo  {journal} {Phys. Rev. X}\ }\textbf {\bibinfo
  {volume} {9}},\ \bibinfo {pages} {021012} (\bibinfo {year}
  {2019})}\BibitemShut {NoStop}%
\bibitem [{\citenamefont {Guo}\ \emph {et~al.}(2019)\citenamefont {Guo},
  \citenamefont {B{\"o}ttcher}, \citenamefont {Hertkorn}, \citenamefont
  {Schmidt}, \citenamefont {Wenzel}, \citenamefont {B{\"u}chler}, \citenamefont
  {Langen},\ and\ \citenamefont {Pfau}}]{guo2019low}%
  \BibitemOpen
  \bibfield  {author} {\bibinfo {author} {\bibfnamefont {M.}~\bibnamefont
  {Guo}}, \bibinfo {author} {\bibfnamefont {F.}~\bibnamefont {B{\"o}ttcher}},
  \bibinfo {author} {\bibfnamefont {J.}~\bibnamefont {Hertkorn}}, \bibinfo
  {author} {\bibfnamefont {J.-N.}\ \bibnamefont {Schmidt}}, \bibinfo {author}
  {\bibfnamefont {M.}~\bibnamefont {Wenzel}}, \bibinfo {author} {\bibfnamefont
  {H.~P.}\ \bibnamefont {B{\"u}chler}}, \bibinfo {author} {\bibfnamefont
  {T.}~\bibnamefont {Langen}}, \ and\ \bibinfo {author} {\bibfnamefont
  {T.}~\bibnamefont {Pfau}},\ }\href {\doibase 10.1038/s41586-019-1569-5}
  {\bibfield  {journal} {\bibinfo  {journal} {Nature}\ }\textbf {\bibinfo
  {volume} {574}},\ \bibinfo {pages} {386} (\bibinfo {year}
  {2019})}\BibitemShut {NoStop}%
\bibitem [{\citenamefont {Natale}\ \emph {et~al.}(2019)\citenamefont {Natale},
  \citenamefont {van Bijnen}, \citenamefont {Patscheider}, \citenamefont
  {Petter}, \citenamefont {Mark}, \citenamefont {Chomaz},\ and\ \citenamefont
  {Ferlaino}}]{PhysRevLett.123.050402}%
  \BibitemOpen
  \bibfield  {author} {\bibinfo {author} {\bibfnamefont {G.}~\bibnamefont
  {Natale}}, \bibinfo {author} {\bibfnamefont {R.~M.~W.}\ \bibnamefont {van
  Bijnen}}, \bibinfo {author} {\bibfnamefont {A.}~\bibnamefont {Patscheider}},
  \bibinfo {author} {\bibfnamefont {D.}~\bibnamefont {Petter}}, \bibinfo
  {author} {\bibfnamefont {M.~J.}\ \bibnamefont {Mark}}, \bibinfo {author}
  {\bibfnamefont {L.}~\bibnamefont {Chomaz}}, \ and\ \bibinfo {author}
  {\bibfnamefont {F.}~\bibnamefont {Ferlaino}},\ }\href {\doibase
  10.1103/PhysRevLett.123.050402} {\bibfield  {journal} {\bibinfo  {journal}
  {Phys. Rev. Lett.}\ }\textbf {\bibinfo {volume} {123}},\ \bibinfo {pages}
  {050402} (\bibinfo {year} {2019})}\BibitemShut {NoStop}%
\bibitem [{\citenamefont {L{\'e}onard}\ \emph {et~al.}(2017)\citenamefont
  {L{\'e}onard}, \citenamefont {Morales}, \citenamefont {Zupancic},
  \citenamefont {Esslinger},\ and\ \citenamefont
  {Donner}}]{leonard2017supersolid}%
  \BibitemOpen
  \bibfield  {author} {\bibinfo {author} {\bibfnamefont {J.}~\bibnamefont
  {L{\'e}onard}}, \bibinfo {author} {\bibfnamefont {A.}~\bibnamefont
  {Morales}}, \bibinfo {author} {\bibfnamefont {P.}~\bibnamefont {Zupancic}},
  \bibinfo {author} {\bibfnamefont {T.}~\bibnamefont {Esslinger}}, \ and\
  \bibinfo {author} {\bibfnamefont {T.}~\bibnamefont {Donner}},\ }\href
  {\doibase 10.1038/nature21067} {\bibfield  {journal} {\bibinfo  {journal}
  {Nature}\ }\textbf {\bibinfo {volume} {543}},\ \bibinfo {pages} {87}
  (\bibinfo {year} {2017})}\BibitemShut {NoStop}%
\bibitem [{\citenamefont {Conti}\ \emph {et~al.}(2023)\citenamefont {Conti},
  \citenamefont {Perali}, \citenamefont {Hamilton}, \citenamefont {Milo\ifmmode
  \check{s}\else \v{s}\fi{}evi\ifmmode~\acute{c}\else \'{c}\fi{}},
  \citenamefont {Peeters},\ and\ \citenamefont
  {Neilson}}]{PhysRevLett.130.057001}%
  \BibitemOpen
  \bibfield  {author} {\bibinfo {author} {\bibfnamefont {S.}~\bibnamefont
  {Conti}}, \bibinfo {author} {\bibfnamefont {A.}~\bibnamefont {Perali}},
  \bibinfo {author} {\bibfnamefont {A.~R.}\ \bibnamefont {Hamilton}}, \bibinfo
  {author} {\bibfnamefont {M.~V.}\ \bibnamefont {Milo\ifmmode \check{s}\else
  \v{s}\fi{}evi\ifmmode~\acute{c}\else \'{c}\fi{}}}, \bibinfo {author}
  {\bibfnamefont {F.~m. c.~M.}\ \bibnamefont {Peeters}}, \ and\ \bibinfo
  {author} {\bibfnamefont {D.}~\bibnamefont {Neilson}},\ }\href {\doibase
  10.1103/PhysRevLett.130.057001} {\bibfield  {journal} {\bibinfo  {journal}
  {Phys. Rev. Lett.}\ }\textbf {\bibinfo {volume} {130}},\ \bibinfo {pages}
  {057001} (\bibinfo {year} {2023})}\BibitemShut {NoStop}%
\bibitem [{\citenamefont {Wang}\ \emph {et~al.}(2010)\citenamefont {Wang},
  \citenamefont {Gao}, \citenamefont {Jian},\ and\ \citenamefont
  {Zhai}}]{PhysRevLett.105.160403}%
  \BibitemOpen
  \bibfield  {author} {\bibinfo {author} {\bibfnamefont {C.}~\bibnamefont
  {Wang}}, \bibinfo {author} {\bibfnamefont {C.}~\bibnamefont {Gao}}, \bibinfo
  {author} {\bibfnamefont {C.-M.}\ \bibnamefont {Jian}}, \ and\ \bibinfo
  {author} {\bibfnamefont {H.}~\bibnamefont {Zhai}},\ }\href {\doibase
  10.1103/PhysRevLett.105.160403} {\bibfield  {journal} {\bibinfo  {journal}
  {Phys. Rev. Lett.}\ }\textbf {\bibinfo {volume} {105}},\ \bibinfo {pages}
  {160403} (\bibinfo {year} {2010})}\BibitemShut {NoStop}%
\bibitem [{\citenamefont {Ho}\ and\ \citenamefont
  {Zhang}(2011)}]{PhysRevLett.107.150403}%
  \BibitemOpen
  \bibfield  {author} {\bibinfo {author} {\bibfnamefont {T.-L.}\ \bibnamefont
  {Ho}}\ and\ \bibinfo {author} {\bibfnamefont {S.}~\bibnamefont {Zhang}},\
  }\href {\doibase 10.1103/PhysRevLett.107.150403} {\bibfield  {journal}
  {\bibinfo  {journal} {Phys. Rev. Lett.}\ }\textbf {\bibinfo {volume} {107}},\
  \bibinfo {pages} {150403} (\bibinfo {year} {2011})}\BibitemShut {NoStop}%
\bibitem [{\citenamefont {Li}\ \emph {et~al.}(2012)\citenamefont {Li},
  \citenamefont {Pitaevskii},\ and\ \citenamefont
  {Stringari}}]{PhysRevLett.108.225301}%
  \BibitemOpen
  \bibfield  {author} {\bibinfo {author} {\bibfnamefont {Y.}~\bibnamefont
  {Li}}, \bibinfo {author} {\bibfnamefont {L.~P.}\ \bibnamefont {Pitaevskii}},
  \ and\ \bibinfo {author} {\bibfnamefont {S.}~\bibnamefont {Stringari}},\
  }\href {\doibase 10.1103/PhysRevLett.108.225301} {\bibfield  {journal}
  {\bibinfo  {journal} {Phys. Rev. Lett.}\ }\textbf {\bibinfo {volume} {108}},\
  \bibinfo {pages} {225301} (\bibinfo {year} {2012})}\BibitemShut {NoStop}%
\bibitem [{\citenamefont {Li}\ \emph {et~al.}(2013)\citenamefont {Li},
  \citenamefont {Martone}, \citenamefont {Pitaevskii},\ and\ \citenamefont
  {Stringari}}]{PhysRevLett.110.235302}%
  \BibitemOpen
  \bibfield  {author} {\bibinfo {author} {\bibfnamefont {Y.}~\bibnamefont
  {Li}}, \bibinfo {author} {\bibfnamefont {G.~I.}\ \bibnamefont {Martone}},
  \bibinfo {author} {\bibfnamefont {L.~P.}\ \bibnamefont {Pitaevskii}}, \ and\
  \bibinfo {author} {\bibfnamefont {S.}~\bibnamefont {Stringari}},\ }\href
  {\doibase 10.1103/PhysRevLett.110.235302} {\bibfield  {journal} {\bibinfo
  {journal} {Phys. Rev. Lett.}\ }\textbf {\bibinfo {volume} {110}},\ \bibinfo
  {pages} {235302} (\bibinfo {year} {2013})}\BibitemShut {NoStop}%
\bibitem [{\citenamefont {Martone}\ \emph {et~al.}(2012)\citenamefont
  {Martone}, \citenamefont {Li}, \citenamefont {Pitaevskii},\ and\
  \citenamefont {Stringari}}]{PhysRevA.86.063621}%
  \BibitemOpen
  \bibfield  {author} {\bibinfo {author} {\bibfnamefont {G.~I.}\ \bibnamefont
  {Martone}}, \bibinfo {author} {\bibfnamefont {Y.}~\bibnamefont {Li}},
  \bibinfo {author} {\bibfnamefont {L.~P.}\ \bibnamefont {Pitaevskii}}, \ and\
  \bibinfo {author} {\bibfnamefont {S.}~\bibnamefont {Stringari}},\ }\href
  {\doibase 10.1103/PhysRevA.86.063621} {\bibfield  {journal} {\bibinfo
  {journal} {Phys. Rev. A}\ }\textbf {\bibinfo {volume} {86}},\ \bibinfo
  {pages} {063621} (\bibinfo {year} {2012})}\BibitemShut {NoStop}%
\bibitem [{\citenamefont {Zheng}\ \emph {et~al.}(2013)\citenamefont {Zheng},
  \citenamefont {Yu}, \citenamefont {Cui},\ and\ \citenamefont
  {Zhai}}]{Zheng_2013}%
  \BibitemOpen
  \bibfield  {author} {\bibinfo {author} {\bibfnamefont {W.}~\bibnamefont
  {Zheng}}, \bibinfo {author} {\bibfnamefont {Z.-Q.}\ \bibnamefont {Yu}},
  \bibinfo {author} {\bibfnamefont {X.}~\bibnamefont {Cui}}, \ and\ \bibinfo
  {author} {\bibfnamefont {H.}~\bibnamefont {Zhai}},\ }\href {\doibase
  10.1088/0953-4075/46/13/134007} {\bibfield  {journal} {\bibinfo  {journal}
  {J. Phys. B}\ }\textbf {\bibinfo {volume} {46}},\ \bibinfo {pages} {134007}
  (\bibinfo {year} {2013})}\BibitemShut {NoStop}%
\bibitem [{\citenamefont {Xia}\ \emph {et~al.}(2023)\citenamefont {Xia},
  \citenamefont {Chen}, \citenamefont {Li}, \citenamefont {Zhang},\ and\
  \citenamefont {Zhu}}]{PhysRevA.107.053302}%
  \BibitemOpen
  \bibfield  {author} {\bibinfo {author} {\bibfnamefont {W.-L.}\ \bibnamefont
  {Xia}}, \bibinfo {author} {\bibfnamefont {L.}~\bibnamefont {Chen}}, \bibinfo
  {author} {\bibfnamefont {T.-T.}\ \bibnamefont {Li}}, \bibinfo {author}
  {\bibfnamefont {Y.}~\bibnamefont {Zhang}}, \ and\ \bibinfo {author}
  {\bibfnamefont {Q.}~\bibnamefont {Zhu}},\ }\href {\doibase
  10.1103/PhysRevA.107.053302} {\bibfield  {journal} {\bibinfo  {journal}
  {Phys. Rev. A}\ }\textbf {\bibinfo {volume} {107}},\ \bibinfo {pages}
  {053302} (\bibinfo {year} {2023})}\BibitemShut {NoStop}%
\bibitem [{\citenamefont {Martone}\ and\ \citenamefont
  {Stringari}(2021)}]{10.21468/SciPostPhys.11.5.092}%
  \BibitemOpen
  \bibfield  {author} {\bibinfo {author} {\bibfnamefont {G.~I.}\ \bibnamefont
  {Martone}}\ and\ \bibinfo {author} {\bibfnamefont {S.}~\bibnamefont
  {Stringari}},\ }\href {\doibase 10.21468/SciPostPhys.11.5.092} {\bibfield
  {journal} {\bibinfo  {journal} {SciPost Phys.}\ }\textbf {\bibinfo {volume}
  {11}},\ \bibinfo {pages} {092} (\bibinfo {year} {2021})}\BibitemShut
  {NoStop}%
\bibitem [{\citenamefont {Leggett}(1970)}]{PhysRevLett.25.1543}%
  \BibitemOpen
  \bibfield  {author} {\bibinfo {author} {\bibfnamefont {A.~J.}\ \bibnamefont
  {Leggett}},\ }\href {\doibase 10.1103/PhysRevLett.25.1543} {\bibfield
  {journal} {\bibinfo  {journal} {Phys. Rev. Lett.}\ }\textbf {\bibinfo
  {volume} {25}},\ \bibinfo {pages} {1543} (\bibinfo {year}
  {1970})}\BibitemShut {NoStop}%
\bibitem [{\citenamefont {Chen}\ \emph
  {et~al.}(2017{\natexlab{a}})\citenamefont {Chen}, \citenamefont {Pu},
  \citenamefont {Yu},\ and\ \citenamefont {Zhang}}]{PhysRevA.95.033616}%
  \BibitemOpen
  \bibfield  {author} {\bibinfo {author} {\bibfnamefont {L.}~\bibnamefont
  {Chen}}, \bibinfo {author} {\bibfnamefont {H.}~\bibnamefont {Pu}}, \bibinfo
  {author} {\bibfnamefont {Z.-Q.}\ \bibnamefont {Yu}}, \ and\ \bibinfo {author}
  {\bibfnamefont {Y.}~\bibnamefont {Zhang}},\ }\href {\doibase
  10.1103/PhysRevA.95.033616} {\bibfield  {journal} {\bibinfo  {journal} {Phys.
  Rev. A}\ }\textbf {\bibinfo {volume} {95}},\ \bibinfo {pages} {033616}
  (\bibinfo {year} {2017}{\natexlab{a}})}\BibitemShut {NoStop}%
\bibitem [{\citenamefont {Geier}\ \emph {et~al.}(2021)\citenamefont {Geier},
  \citenamefont {Martone}, \citenamefont {Hauke},\ and\ \citenamefont
  {Stringari}}]{PhysRevLett.127.115301}%
  \BibitemOpen
  \bibfield  {author} {\bibinfo {author} {\bibfnamefont {K.~T.}\ \bibnamefont
  {Geier}}, \bibinfo {author} {\bibfnamefont {G.~I.}\ \bibnamefont {Martone}},
  \bibinfo {author} {\bibfnamefont {P.}~\bibnamefont {Hauke}}, \ and\ \bibinfo
  {author} {\bibfnamefont {S.}~\bibnamefont {Stringari}},\ }\href {\doibase
  10.1103/PhysRevLett.127.115301} {\bibfield  {journal} {\bibinfo  {journal}
  {Phys. Rev. Lett.}\ }\textbf {\bibinfo {volume} {127}},\ \bibinfo {pages}
  {115301} (\bibinfo {year} {2021})}\BibitemShut {NoStop}%
\bibitem [{\citenamefont {Geier}\ \emph {et~al.}(2023)\citenamefont {Geier},
  \citenamefont {Martone}, \citenamefont {Hauke}, \citenamefont {Ketterle},\
  and\ \citenamefont {Stringari}}]{PhysRevLett.130.156001}%
  \BibitemOpen
  \bibfield  {author} {\bibinfo {author} {\bibfnamefont {K.~T.}\ \bibnamefont
  {Geier}}, \bibinfo {author} {\bibfnamefont {G.~I.}\ \bibnamefont {Martone}},
  \bibinfo {author} {\bibfnamefont {P.}~\bibnamefont {Hauke}}, \bibinfo
  {author} {\bibfnamefont {W.}~\bibnamefont {Ketterle}}, \ and\ \bibinfo
  {author} {\bibfnamefont {S.}~\bibnamefont {Stringari}},\ }\href {\doibase
  10.1103/PhysRevLett.130.156001} {\bibfield  {journal} {\bibinfo  {journal}
  {Phys. Rev. Lett.}\ }\textbf {\bibinfo {volume} {130}},\ \bibinfo {pages}
  {156001} (\bibinfo {year} {2023})}\BibitemShut {NoStop}%
\bibitem [{\citenamefont {Chen}\ \emph {et~al.}(2022)\citenamefont {Chen},
  \citenamefont {Liu},\ and\ \citenamefont {Hu}}]{PhysRevA.106.023302}%
  \BibitemOpen
  \bibfield  {author} {\bibinfo {author} {\bibfnamefont {X.-L.}\ \bibnamefont
  {Chen}}, \bibinfo {author} {\bibfnamefont {X.-J.}\ \bibnamefont {Liu}}, \
  and\ \bibinfo {author} {\bibfnamefont {H.}~\bibnamefont {Hu}},\ }\href
  {\doibase 10.1103/PhysRevA.106.023302} {\bibfield  {journal} {\bibinfo
  {journal} {Phys. Rev. A}\ }\textbf {\bibinfo {volume} {106}},\ \bibinfo
  {pages} {023302} (\bibinfo {year} {2022})}\BibitemShut {NoStop}%
\bibitem [{\citenamefont {Chen}\ \emph
  {et~al.}(2017{\natexlab{b}})\citenamefont {Chen}, \citenamefont {Liu},\ and\
  \citenamefont {Hu}}]{PhysRevA.96.013625}%
  \BibitemOpen
  \bibfield  {author} {\bibinfo {author} {\bibfnamefont {X.-L.}\ \bibnamefont
  {Chen}}, \bibinfo {author} {\bibfnamefont {X.-J.}\ \bibnamefont {Liu}}, \
  and\ \bibinfo {author} {\bibfnamefont {H.}~\bibnamefont {Hu}},\ }\href
  {\doibase 10.1103/PhysRevA.96.013625} {\bibfield  {journal} {\bibinfo
  {journal} {Phys. Rev. A}\ }\textbf {\bibinfo {volume} {96}},\ \bibinfo
  {pages} {013625} (\bibinfo {year} {2017}{\natexlab{b}})}\BibitemShut
  {NoStop}%
\bibitem [{\citenamefont {Yu}(2014)}]{PhysRevA.90.053608}%
  \BibitemOpen
  \bibfield  {author} {\bibinfo {author} {\bibfnamefont {Z.-Q.}\ \bibnamefont
  {Yu}},\ }\href {\doibase 10.1103/PhysRevA.90.053608} {\bibfield  {journal}
  {\bibinfo  {journal} {Phys. Rev. A}\ }\textbf {\bibinfo {volume} {90}},\
  \bibinfo {pages} {053608} (\bibinfo {year} {2014})}\BibitemShut {NoStop}%
\bibitem [{\citenamefont {Griffin}(1996)}]{PhysRevB.53.9341}%
  \BibitemOpen
  \bibfield  {author} {\bibinfo {author} {\bibfnamefont {A.}~\bibnamefont
  {Griffin}},\ }\href {\doibase 10.1103/PhysRevB.53.9341} {\bibfield  {journal}
  {\bibinfo  {journal} {Phys. Rev. B}\ }\textbf {\bibinfo {volume} {53}},\
  \bibinfo {pages} {9341} (\bibinfo {year} {1996})}\BibitemShut {NoStop}%
\bibitem [{\citenamefont {Lin}\ \emph {et~al.}(2011)\citenamefont {Lin},
  \citenamefont {Jim{\'e}nez-Garc{\'\i}a},\ and\ \citenamefont
  {Spielman}}]{Lin2011}%
  \BibitemOpen
  \bibfield  {author} {\bibinfo {author} {\bibfnamefont {Y.-J.}\ \bibnamefont
  {Lin}}, \bibinfo {author} {\bibfnamefont {K.}~\bibnamefont
  {Jim{\'e}nez-Garc{\'\i}a}}, \ and\ \bibinfo {author} {\bibfnamefont {I.~B.}\
  \bibnamefont {Spielman}},\ }\href {\doibase 10.1038/nature09887} {\bibfield
  {journal} {\bibinfo  {journal} {Nature (London)}\ }\textbf {\bibinfo {volume}
  {471}},\ \bibinfo {pages} {83} (\bibinfo {year} {2011})}\BibitemShut
  {NoStop}%
\bibitem [{\citenamefont {Roy}\ \emph {et~al.}(2014)\citenamefont {Roy},
  \citenamefont {Gautam},\ and\ \citenamefont {Angom}}]{PhysRevA.89.013617}%
  \BibitemOpen
  \bibfield  {author} {\bibinfo {author} {\bibfnamefont {A.}~\bibnamefont
  {Roy}}, \bibinfo {author} {\bibfnamefont {S.}~\bibnamefont {Gautam}}, \ and\
  \bibinfo {author} {\bibfnamefont {D.}~\bibnamefont {Angom}},\ }\href
  {\doibase 10.1103/PhysRevA.89.013617} {\bibfield  {journal} {\bibinfo
  {journal} {Phys. Rev. A}\ }\textbf {\bibinfo {volume} {89}},\ \bibinfo
  {pages} {013617} (\bibinfo {year} {2014})}\BibitemShut {NoStop}%
\bibitem [{\citenamefont {Rajat}\ \emph {et~al.}(2022)\citenamefont {Rajat},
  \citenamefont {Roy},\ and\ \citenamefont {Gautam}}]{PhysRevA.106.013304}%
  \BibitemOpen
  \bibfield  {author} {\bibinfo {author} {\bibnamefont {Rajat}}, \bibinfo
  {author} {\bibfnamefont {A.}~\bibnamefont {Roy}}, \ and\ \bibinfo {author}
  {\bibfnamefont {S.}~\bibnamefont {Gautam}},\ }\href {\doibase
  10.1103/PhysRevA.106.013304} {\bibfield  {journal} {\bibinfo  {journal}
  {Phys. Rev. A}\ }\textbf {\bibinfo {volume} {106}},\ \bibinfo {pages}
  {013304} (\bibinfo {year} {2022})}\BibitemShut {NoStop}%
\bibitem [{\citenamefont {Popov}(1987)}]{popov1987functional}%
  \BibitemOpen
  \bibfield  {author} {\bibinfo {author} {\bibfnamefont {V.~N.}\ \bibnamefont
  {Popov}},\ }\href@noop {} {\emph {\bibinfo {title} {Functional integrals and
  collective excitations}}}\ (\bibinfo  {publisher} {Cambridge University
  Press},\ \bibinfo {year} {1987})\BibitemShut {NoStop}%
\bibitem [{\citenamefont {Kaur}\ \emph {et~al.}(2021)\citenamefont {Kaur},
  \citenamefont {Roy},\ and\ \citenamefont {Gautam}}]{kaur2021fortress}%
  \BibitemOpen
  \bibfield  {author} {\bibinfo {author} {\bibfnamefont {P.}~\bibnamefont
  {Kaur}}, \bibinfo {author} {\bibfnamefont {A.}~\bibnamefont {Roy}}, \ and\
  \bibinfo {author} {\bibfnamefont {S.}~\bibnamefont {Gautam}},\ }\href
  {\doibase https://doi.org/10.1016/j.cpc.2020.107671} {\bibfield  {journal}
  {\bibinfo  {journal} {Comput. Phys. Commun.}\ }\textbf {\bibinfo {volume}
  {259}},\ \bibinfo {pages} {107671} (\bibinfo {year} {2021})}\BibitemShut
  {NoStop}%
\bibitem [{\citenamefont {Banger}\ \emph
  {et~al.}(2022{\natexlab{a}})\citenamefont {Banger}, \citenamefont {Kaur},\
  and\ \citenamefont {Gautam}}]{banger2021semi}%
  \BibitemOpen
  \bibfield  {author} {\bibinfo {author} {\bibfnamefont {P.}~\bibnamefont
  {Banger}}, \bibinfo {author} {\bibfnamefont {P.}~\bibnamefont {Kaur}}, \ and\
  \bibinfo {author} {\bibfnamefont {S.}~\bibnamefont {Gautam}},\ }\href
  {\doibase 10.1142/S0129183122500462} {\bibfield  {journal} {\bibinfo
  {journal} {Int. J. Mod. Phys. C}\ }\textbf {\bibinfo {volume} {33}},\
  \bibinfo {pages} {2250046} (\bibinfo {year}
  {2022}{\natexlab{a}})}\BibitemShut {NoStop}%
\bibitem [{\citenamefont {Banger}\ \emph
  {et~al.}(2022{\natexlab{b}})\citenamefont {Banger}, \citenamefont {Kaur},
  \citenamefont {Roy},\ and\ \citenamefont {Gautam}}]{banger2022fortress}%
  \BibitemOpen
  \bibfield  {author} {\bibinfo {author} {\bibfnamefont {P.}~\bibnamefont
  {Banger}}, \bibinfo {author} {\bibfnamefont {P.}~\bibnamefont {Kaur}},
  \bibinfo {author} {\bibfnamefont {A.}~\bibnamefont {Roy}}, \ and\ \bibinfo
  {author} {\bibfnamefont {S.}~\bibnamefont {Gautam}},\ }\href {\doibase
  https://doi.org/10.1016/j.cpc.2022.108442} {\bibfield  {journal} {\bibinfo
  {journal} {Comput. Phys. Commun.}\ }\textbf {\bibinfo {volume} {279}},\
  \bibinfo {pages} {108442} (\bibinfo {year} {2022}{\natexlab{b}})}\BibitemShut
  {NoStop}%
\bibitem [{\citenamefont {Dodd}\ \emph {et~al.}(1998)\citenamefont {Dodd},
  \citenamefont {Edwards}, \citenamefont {Clark},\ and\ \citenamefont
  {Burnett}}]{PhysRevA.57.R32}%
  \BibitemOpen
  \bibfield  {author} {\bibinfo {author} {\bibfnamefont {R.~J.}\ \bibnamefont
  {Dodd}}, \bibinfo {author} {\bibfnamefont {M.}~\bibnamefont {Edwards}},
  \bibinfo {author} {\bibfnamefont {C.~W.}\ \bibnamefont {Clark}}, \ and\
  \bibinfo {author} {\bibfnamefont {K.}~\bibnamefont {Burnett}},\ }\href
  {\doibase 10.1103/PhysRevA.57.R32} {\bibfield  {journal} {\bibinfo  {journal}
  {Phys. Rev. A}\ }\textbf {\bibinfo {volume} {57}},\ \bibinfo {pages} {R32}
  (\bibinfo {year} {1998})}\BibitemShut {NoStop}%
\bibitem [{\citenamefont {Roy}\ \emph {et~al.}(2020)\citenamefont {Roy},
  \citenamefont {Pal}, \citenamefont {Gautam}, \citenamefont {Angom},\ and\
  \citenamefont {Muruganandam}}]{roy-2020}%
  \BibitemOpen
  \bibfield  {author} {\bibinfo {author} {\bibfnamefont {A.}~\bibnamefont
  {Roy}}, \bibinfo {author} {\bibfnamefont {S.}~\bibnamefont {Pal}}, \bibinfo
  {author} {\bibfnamefont {S.}~\bibnamefont {Gautam}}, \bibinfo {author}
  {\bibfnamefont {D.}~\bibnamefont {Angom}}, \ and\ \bibinfo {author}
  {\bibfnamefont {P.}~\bibnamefont {Muruganandam}},\ }\href {\doibase
  https://doi.org/10.1016/j.cpc.2020.107288} {\bibfield  {journal} {\bibinfo
  {journal} {Comput. Phys. Commun.}\ }\textbf {\bibinfo {volume} {256}},\
  \bibinfo {pages} {107288} (\bibinfo {year} {2020})}\BibitemShut {NoStop}%
\bibitem [{\citenamefont {Knoop}\ \emph {et~al.}(2011)\citenamefont {Knoop},
  \citenamefont {Schuster}, \citenamefont {Scelle}, \citenamefont {Trautmann},
  \citenamefont {Appmeier}, \citenamefont {Oberthaler}, \citenamefont
  {Tiesinga},\ and\ \citenamefont {Tiemann}}]{PhysRevA.83.042704}%
  \BibitemOpen
  \bibfield  {author} {\bibinfo {author} {\bibfnamefont {S.}~\bibnamefont
  {Knoop}}, \bibinfo {author} {\bibfnamefont {T.}~\bibnamefont {Schuster}},
  \bibinfo {author} {\bibfnamefont {R.}~\bibnamefont {Scelle}}, \bibinfo
  {author} {\bibfnamefont {A.}~\bibnamefont {Trautmann}}, \bibinfo {author}
  {\bibfnamefont {J.}~\bibnamefont {Appmeier}}, \bibinfo {author}
  {\bibfnamefont {M.~K.}\ \bibnamefont {Oberthaler}}, \bibinfo {author}
  {\bibfnamefont {E.}~\bibnamefont {Tiesinga}}, \ and\ \bibinfo {author}
  {\bibfnamefont {E.}~\bibnamefont {Tiemann}},\ }\href {\doibase
  10.1103/PhysRevA.83.042704} {\bibfield  {journal} {\bibinfo  {journal} {Phys.
  Rev. A}\ }\textbf {\bibinfo {volume} {83}},\ \bibinfo {pages} {042704}
  (\bibinfo {year} {2011})}\BibitemShut {NoStop}%
\bibitem [{\citenamefont {Martone}\ \emph {et~al.}(2014)\citenamefont
  {Martone}, \citenamefont {Li},\ and\ \citenamefont
  {Stringari}}]{PhysRevA.90.041604}%
  \BibitemOpen
  \bibfield  {author} {\bibinfo {author} {\bibfnamefont {G.~I.}\ \bibnamefont
  {Martone}}, \bibinfo {author} {\bibfnamefont {Y.}~\bibnamefont {Li}}, \ and\
  \bibinfo {author} {\bibfnamefont {S.}~\bibnamefont {Stringari}},\ }\href
  {\doibase 10.1103/PhysRevA.90.041604} {\bibfield  {journal} {\bibinfo
  {journal} {Phys. Rev. A}\ }\textbf {\bibinfo {volume} {90}},\ \bibinfo
  {pages} {041604} (\bibinfo {year} {2014})}\BibitemShut {NoStop}%
\bibitem [{\citenamefont {Ketterle}\ and\ \citenamefont {van
  Druten}(1996)}]{PhysRevA.54.656}%
  \BibitemOpen
  \bibfield  {author} {\bibinfo {author} {\bibfnamefont {W.}~\bibnamefont
  {Ketterle}}\ and\ \bibinfo {author} {\bibfnamefont {N.~J.}\ \bibnamefont {van
  Druten}},\ }\href {\doibase 10.1103/PhysRevA.54.656} {\bibfield  {journal}
  {\bibinfo  {journal} {Phys. Rev. A}\ }\textbf {\bibinfo {volume} {54}},\
  \bibinfo {pages} {656} (\bibinfo {year} {1996})}\BibitemShut {NoStop}%
\bibitem [{\citenamefont {Ji}\ \emph {et~al.}(2014)\citenamefont {Ji},
  \citenamefont {Zhang}, \citenamefont {Zhang}, \citenamefont {Du},
  \citenamefont {Zheng}, \citenamefont {Deng}, \citenamefont {Zhai},
  \citenamefont {Chen},\ and\ \citenamefont {Pan}}]{ji2014experimental}%
  \BibitemOpen
  \bibfield  {author} {\bibinfo {author} {\bibfnamefont {S.-C.}\ \bibnamefont
  {Ji}}, \bibinfo {author} {\bibfnamefont {J.-Y.}\ \bibnamefont {Zhang}},
  \bibinfo {author} {\bibfnamefont {L.}~\bibnamefont {Zhang}}, \bibinfo
  {author} {\bibfnamefont {Z.-D.}\ \bibnamefont {Du}}, \bibinfo {author}
  {\bibfnamefont {W.}~\bibnamefont {Zheng}}, \bibinfo {author} {\bibfnamefont
  {Y.-J.}\ \bibnamefont {Deng}}, \bibinfo {author} {\bibfnamefont
  {H.}~\bibnamefont {Zhai}}, \bibinfo {author} {\bibfnamefont {S.}~\bibnamefont
  {Chen}}, \ and\ \bibinfo {author} {\bibfnamefont {J.-W.}\ \bibnamefont
  {Pan}},\ }\href {\doibase 10.1038/nphys2905} {\bibfield  {journal} {\bibinfo
  {journal} {Nature physics}\ }\textbf {\bibinfo {volume} {10}},\ \bibinfo
  {pages} {314} (\bibinfo {year} {2014})}\BibitemShut {NoStop}%
\bibitem [{\citenamefont {Liao}\ \emph {et~al.}(2014)\citenamefont {Liao},
  \citenamefont {Huang}, \citenamefont {Lin},\ and\ \citenamefont
  {Fialko}}]{PhysRevA.89.063614}%
  \BibitemOpen
  \bibfield  {author} {\bibinfo {author} {\bibfnamefont {R.}~\bibnamefont
  {Liao}}, \bibinfo {author} {\bibfnamefont {Z.-G.}\ \bibnamefont {Huang}},
  \bibinfo {author} {\bibfnamefont {X.-M.}\ \bibnamefont {Lin}}, \ and\
  \bibinfo {author} {\bibfnamefont {O.}~\bibnamefont {Fialko}},\ }\href
  {\doibase 10.1103/PhysRevA.89.063614} {\bibfield  {journal} {\bibinfo
  {journal} {Phys. Rev. A}\ }\textbf {\bibinfo {volume} {89}},\ \bibinfo
  {pages} {063614} (\bibinfo {year} {2014})}\BibitemShut {NoStop}%
\bibitem [{\citenamefont {Banger}\ \emph {et~al.}(2023)\citenamefont {Banger},
  \citenamefont {Rajat}, \citenamefont {Roy},\ and\ \citenamefont
  {Gautam}}]{PhysRevA.108.043310}%
  \BibitemOpen
  \bibfield  {author} {\bibinfo {author} {\bibfnamefont {P.}~\bibnamefont
  {Banger}}, \bibinfo {author} {\bibnamefont {Rajat}}, \bibinfo {author}
  {\bibfnamefont {A.}~\bibnamefont {Roy}}, \ and\ \bibinfo {author}
  {\bibfnamefont {S.}~\bibnamefont {Gautam}},\ }\href {\doibase
  10.1103/PhysRevA.108.043310} {\bibfield  {journal} {\bibinfo  {journal}
  {Phys. Rev. A}\ }\textbf {\bibinfo {volume} {108}},\ \bibinfo {pages}
  {043310} (\bibinfo {year} {2023})}\BibitemShut {NoStop}%
\bibitem [{\citenamefont {Wilson}\ \emph {et~al.}(2010)\citenamefont {Wilson},
  \citenamefont {Ronen},\ and\ \citenamefont {Bohn}}]{PhysRevLett.104.094501}%
  \BibitemOpen
  \bibfield  {author} {\bibinfo {author} {\bibfnamefont {R.~M.}\ \bibnamefont
  {Wilson}}, \bibinfo {author} {\bibfnamefont {S.}~\bibnamefont {Ronen}}, \
  and\ \bibinfo {author} {\bibfnamefont {J.~L.}\ \bibnamefont {Bohn}},\ }\href
  {\doibase 10.1103/PhysRevLett.104.094501} {\bibfield  {journal} {\bibinfo
  {journal} {Phys. Rev. Lett.}\ }\textbf {\bibinfo {volume} {104}},\ \bibinfo
  {pages} {094501} (\bibinfo {year} {2010})}\BibitemShut {NoStop}%
\bibitem [{\citenamefont {Khamehchi}\ \emph {et~al.}(2014)\citenamefont
  {Khamehchi}, \citenamefont {Zhang}, \citenamefont {Hamner}, \citenamefont
  {Busch},\ and\ \citenamefont {Engels}}]{PhysRevA.90.063624}%
  \BibitemOpen
  \bibfield  {author} {\bibinfo {author} {\bibfnamefont {M.~A.}\ \bibnamefont
  {Khamehchi}}, \bibinfo {author} {\bibfnamefont {Y.}~\bibnamefont {Zhang}},
  \bibinfo {author} {\bibfnamefont {C.}~\bibnamefont {Hamner}}, \bibinfo
  {author} {\bibfnamefont {T.}~\bibnamefont {Busch}}, \ and\ \bibinfo {author}
  {\bibfnamefont {P.}~\bibnamefont {Engels}},\ }\href {\doibase
  10.1103/PhysRevA.90.063624} {\bibfield  {journal} {\bibinfo  {journal} {Phys.
  Rev. A}\ }\textbf {\bibinfo {volume} {90}},\ \bibinfo {pages} {063624}
  (\bibinfo {year} {2014})}\BibitemShut {NoStop}%
\bibitem [{\citenamefont {Ji}\ \emph {et~al.}(2015)\citenamefont {Ji},
  \citenamefont {Zhang}, \citenamefont {Xu}, \citenamefont {Wu}, \citenamefont
  {Deng}, \citenamefont {Chen},\ and\ \citenamefont
  {Pan}}]{PhysRevLett.114.105301}%
  \BibitemOpen
  \bibfield  {author} {\bibinfo {author} {\bibfnamefont {S.-C.}\ \bibnamefont
  {Ji}}, \bibinfo {author} {\bibfnamefont {L.}~\bibnamefont {Zhang}}, \bibinfo
  {author} {\bibfnamefont {X.-T.}\ \bibnamefont {Xu}}, \bibinfo {author}
  {\bibfnamefont {Z.}~\bibnamefont {Wu}}, \bibinfo {author} {\bibfnamefont
  {Y.}~\bibnamefont {Deng}}, \bibinfo {author} {\bibfnamefont {S.}~\bibnamefont
  {Chen}}, \ and\ \bibinfo {author} {\bibfnamefont {J.-W.}\ \bibnamefont
  {Pan}},\ }\href {\doibase 10.1103/PhysRevLett.114.105301} {\bibfield
  {journal} {\bibinfo  {journal} {Phys. Rev. Lett.}\ }\textbf {\bibinfo
  {volume} {114}},\ \bibinfo {pages} {105301} (\bibinfo {year}
  {2015})}\BibitemShut {NoStop}%
\bibitem [{\citenamefont {Bisset}\ \emph {et~al.}(2013)\citenamefont {Bisset},
  \citenamefont {Baillie},\ and\ \citenamefont {Blakie}}]{PhysRevA.88.043606}%
  \BibitemOpen
  \bibfield  {author} {\bibinfo {author} {\bibfnamefont {R.~N.}\ \bibnamefont
  {Bisset}}, \bibinfo {author} {\bibfnamefont {D.}~\bibnamefont {Baillie}}, \
  and\ \bibinfo {author} {\bibfnamefont {P.~B.}\ \bibnamefont {Blakie}},\
  }\href {\doibase 10.1103/PhysRevA.88.043606} {\bibfield  {journal} {\bibinfo
  {journal} {Phys. Rev. A}\ }\textbf {\bibinfo {volume} {88}},\ \bibinfo
  {pages} {043606} (\bibinfo {year} {2013})}\BibitemShut {NoStop}%
\bibitem [{\citenamefont {Sohmen}\ \emph {et~al.}(2021)\citenamefont {Sohmen},
  \citenamefont {Politi}, \citenamefont {Klaus}, \citenamefont {Chomaz},
  \citenamefont {Mark}, \citenamefont {Norcia},\ and\ \citenamefont
  {Ferlaino}}]{PhysRevLett.126.233401}%
  \BibitemOpen
  \bibfield  {author} {\bibinfo {author} {\bibfnamefont {M.}~\bibnamefont
  {Sohmen}}, \bibinfo {author} {\bibfnamefont {C.}~\bibnamefont {Politi}},
  \bibinfo {author} {\bibfnamefont {L.}~\bibnamefont {Klaus}}, \bibinfo
  {author} {\bibfnamefont {L.}~\bibnamefont {Chomaz}}, \bibinfo {author}
  {\bibfnamefont {M.~J.}\ \bibnamefont {Mark}}, \bibinfo {author}
  {\bibfnamefont {M.~A.}\ \bibnamefont {Norcia}}, \ and\ \bibinfo {author}
  {\bibfnamefont {F.}~\bibnamefont {Ferlaino}},\ }\href {\doibase
  10.1103/PhysRevLett.126.233401} {\bibfield  {journal} {\bibinfo  {journal}
  {Phys. Rev. Lett.}\ }\textbf {\bibinfo {volume} {126}},\ \bibinfo {pages}
  {233401} (\bibinfo {year} {2021})}\BibitemShut {NoStop}%
\bibitem [{\citenamefont {S{\'a}nchez-Baena}\ \emph {et~al.}(2023)\citenamefont
  {S{\'a}nchez-Baena}, \citenamefont {Politi}, \citenamefont {Maucher},
  \citenamefont {Ferlaino},\ and\ \citenamefont {Pohl}}]{sanchez2023heating}%
  \BibitemOpen
  \bibfield  {author} {\bibinfo {author} {\bibfnamefont {J.}~\bibnamefont
  {S{\'a}nchez-Baena}}, \bibinfo {author} {\bibfnamefont {C.}~\bibnamefont
  {Politi}}, \bibinfo {author} {\bibfnamefont {F.}~\bibnamefont {Maucher}},
  \bibinfo {author} {\bibfnamefont {F.}~\bibnamefont {Ferlaino}}, \ and\
  \bibinfo {author} {\bibfnamefont {T.}~\bibnamefont {Pohl}},\ }\href {\doibase
  10.1038/s41467-023-37207-3} {\bibfield  {journal} {\bibinfo  {journal} {Nat.
  Commun.}\ }\textbf {\bibinfo {volume} {14}},\ \bibinfo {pages} {1868}
  (\bibinfo {year} {2023})}\BibitemShut {NoStop}%
\bibitem [{\citenamefont {Radzihovsky}()}]{radzihovsky2023reentrant}%
  \BibitemOpen
  \bibfield  {author} {\bibinfo {author} {\bibfnamefont {L.}~\bibnamefont
  {Radzihovsky}},\ }\href {https://doi.org/10.48550/arXiv.2311.04266} {\bibinfo
   {journal} {arXiv:2311.04266}\ }\BibitemShut {NoStop}%
\end{thebibliography}%
\bibliographystyle{apsrev4-1}
\end{document}